\documentstyle[11pt,paspconf,twoside,galdyn,psfig]{article}

\begin{document}
\title{Galaxy Formation, Bars and QSOs}
\author{J. A. Sellwood}
\affil{Department of Physics \& Astronomy, Rutgers University, \\
136 Frelinghuysen Road, Piscataway, NJ 08854, USA}

\begin{abstract}
A model that accounts for the brief flaring of QSOs in the early stages of galaxy formation is proposed.  I argue that a bar must develop early in the life of nearly every galaxy and that gas to create and fuel the QSO is driven into the center of the galaxy by the bar.  The QSO lifetime, and the mass of its central engine, are also controlled by large-scale dynamics, since the fuel supply is shut off after a short period by the development of an inner Lindblad resonance.  This resonance causes the gas inflow along the bar to stall at a distance of a few hundred parsecs from the center.  The ILR develops as a result of previous inflow, making quasar activity self-limiting.  The bars are weakened and can be destroyed by the central mass concentration formed in this way.
\end{abstract}

\section{Introduction}
It is often noted that bright galaxies were assembled at about the same time that QSOs flare (\eg\ Rees 1997), suggesting a causal connection.  Blandford (this volume) reviews both the evidence for short lifetimes for QSOs and some recent models for their formation, but leaves open the question of whether they formed before or after the first galaxies.  I suggest that, since QSOs are believed to reside in the centers of galaxies (\eg\ Bahcall \etal\ 1997), it is natural to suppose that they formed there.  There may be a loose proportionality between the mass of the BH and that of the bulge in which it resides (Magorrian \etal\ 1998).  Thus an attractive model would offer answers to at least the following questions:
\begin{itemize}
\item Why should QSOs flare during an early stage of galaxy formation?
\item Why are the centers of galaxies the preferred sites for QSOs?
\item What interrupts the fuel supply to limit QSO lifetimes?
\item Why is the mass of the central BH related to that of the host galaxy?
\end{itemize}
Here I outline a model that offers dynamical answers to all these questions.  The main ideas are: (1) most large galaxies developed a bar at an early stage of their formation, (2) the central engine is created from gas driven to the center by the bar and (3) changes to the galaxy potential, caused by mass inflow itself, shut off the fuel supply to the central engine when the mass concentration reaches a small fraction of the galaxy mass.  Furthermore, this central mass is sufficient to weaken or even destroy the bar. 

\section{Bars in young galaxies}
I adopt the conventional picture that a galaxy disk forms as gas cools and settles into rotational balance in a dark matter halo.  As I argue elsewhere (Sellwood, this volume), the DM halo has a large, low-density core.  Unless the cooling gas has very low angular momentum, the disk it forms will be extensive and have a gently rising rotation curve at first.  Under these conditions, a global bar instability will become unavoidable as the mass of the disk rises (\eg\ Binney \& Tremaine 1987, \S6.3).  Thus almost every galaxy with a dominant disk today would have become barred early in its lifetime.

\section{Gas inflow driven by bars}
Many authors (\eg\ Shlosman \etal\ 1990) have suggested that bar-driven gas inflow could fuel QSO activity.  The inflow occurs because large-scale shocks develop in the gas flow pattern within the bar which Prendergast (1962) identified with the dust lanes seen along the leading sides of bars in galaxies today.  The gas loses both energy and angular momentum in these shocks, the latter because the gas is asymmetrically distributed about the bar major axis.  Thus gas is driven towards the center.

Of the many simulations of gas flows in barred galaxy-like potentials, those by Athanassoula (1992) perhaps illustrate most clearly the difference in flow patterns caused by a central mass.  A relatively shallow density profile allows gas to flow right into the center (her Figure 4), whereas a mass concentration causes the gas flow to stall some distance from the center (her Figure 2).  The different flow pattern in the second case results from the presence of a generalized inner Lindblad resonance of the bar; outside this resonance the flow pattern is generally aligned along the bar, but it switches to perpendicular alignment inside this radius.  The flow pattern inside this resonance ring does not contain shocks, and the gas cannot be driven by the bar any closer to the center.  This phenomenon is also seen in nearby barred galaxies: nuclear rings occur at radii of a few hundred parsecs in many barred galaxies (Buta \& Crocker 1993) where gas is often observed to pile up (Helfer \& Blitz 1995; Rubin \etal\ 1997).

\section{Quasar activity}
The bar which forms early in the life of a galaxy lacks a central mass concentration and does not possess an ILR.  The abundant gas at this epoch will therefore be driven close to the center by the bar.  But as the mass in the center rises to a percent or two of the then galaxy mass, an ILR will develop shutting off the dynamically driven flow of gas into the very inner regions.  Thus the amount of gas that can reach the central $\sim 50$ pc is naturally limited by the large-scale dynamics.

It is hard to predict the precise fate of the gas as it accumulates in the center, but an attractive guess is that some fraction of it makes a collapsed object while the rest forms stars.

As bars form on the dynamical time-scale of the inner galaxy, and gas inflow time is not much longer, we expect a central engine to be created soon after a galaxy begins to be assembled.  The ILR valve will close shortly thereafter, depriving the central collapsed object of further fuel and limiting its mass to a small fraction of the galactic mass.

\section{Bar destruction}
The majority of galaxies are not strongly barred today, so the above picture requires that most bars be destroyed.  Simulators have been reporting for years that stellar bars seem to be robust, long-lived objects (\eg\ Miller \& Smith 1979; Sparke \& Sellwood 1986), but it is now known that bars can be destroyed by growing a central object.  The mass and concentration needed for a central object to destroy a bar is not known at all precisely; Norman \etal\ (1996) showed that a dense object containing 5\% of the disk plus bulge mass caused the bar to dissolve on a dynamical time, but Friedli's (1994) work suggests that masses of 1-2\% could lead to slower bar decay (see also Hozumi \& Hernquist 1998).  The central masses needed seem too high to be just the collapsed object, but all the gas and stars within a radius $\sim 50$ pc should be included.

Sellwood \& Moore (1999) report simulations that mimic this process.  They grow a central mass ``by hand'' after a bar develops and limit its mass to 1.5\% of the initial galaxy mass, to mimic the effect of the formation of an ILR.  They find that the bar is weakened at this stage, though not yet totally destroyed.

They go on to mimic later infall of fresh material which causes strong spiral patterns to develop in the disk.  In some cases, the spiral patterns are vigorous enough to destroy the already weakened bar, but in other cases, infall of the material with the appropriate angular momentum can re-excite the bar.

\section{Conclusions}
I have argued that every bright galaxy should have developed a bar early in its lifetime.  The bar drives gas into the center which creates (in an unspecified manner) a central engine for QSO activity.  Once the collapsed object, and its surrounding gas and star cluster, reaches a mass of 1-2\% of the (luminous) galaxy mass at that time, an inner Lindblad resonance forms which shuts off the dynamically driven gas supply to the central engine.  The dense center also weakens the bar, which may either be destroyed or re-excited, depending on the angular momentum distribution of later infalling material.

This model leads two significant predictions: (1) Halo dominated galaxies, such as LSB or low-luminosity galaxies, which never suffered a bar instability, should not contain massive BHs.  A good example of the latter kind is M33, for which Kormendy \& McClure (1993) have placed a very low upper limit of $10^4\;$\Msun\ on the mass of any central BH.  (2) The fraction of barred galaxies should be {\it lower\/} in the early universe.  This is because the first barred phase should be very short and occurs when the QSO is bright making the bar hard to see.  By the time the QSO fades, any residual bar will be short and weak, and the later development of large-scale bars in galaxies is a more gradual process.  Some observational support for this prediction is now available (Abraham \etal\ 1998).

The model proposed here does not exclude the possibility that QSO activity would be re-ignited during galaxy mergers.  Indeed, the further growth of the BHs during/after a merger will lead to brighter QSOs than those expected in the early stages because the central engines will be more massive.

\acknowledgments
This work was supported by NSF grant AST 96/17088 and NASA LTSA grant NAG 5-6037.

\end{document}